# From Resource Control to Digital Trust with User-Managed Access

White paper – SolidLab[1] project
Version: 2024-11-05

*W. Termont[2], R. Dedecker[3], W. Slabbinck[4],
B. Esteves[5], B. De Meester[6], and R. Verborgh[7]*

*IDLab[8], Ghent University – imec*

---

[1] https://solidlab.be
[2] Wouter Termont, wouter.termont@ugent.be
[3] Ruben Dedecker, ruben.dedecker@ugent.be
[4] Wout Slabbinck, wout.slabbinck@ugent.be
[5] Dr. ir. Beatriz Esteves, beatriz.esteves@ugent.be
[6] Dr. ir. Ben De Meester, ben.demeester@ugent.be
[7] Prof. dr. ir. Ruben Verborgh, ruben.verborgh@ugent.be
[8] https://idlab.ugent.be

# Executive summary

*The User-Managed Access (UMA) extension to OAuth 2.0 is a promising candidate for increasing Digital Trust in personal data ecosystems like Solid. With minor modifications, it can achieve many requirements regarding usage control and transaction contextualization, even though additional specification is needed to address delegation of control and retraction of usage policies.*

# Abstract

*Use cases involving private resources are missing out on the benefits of the Web of Data, and remain stuck in a fundamental lack of Digital Trust. The nature of traditional access control mechanisms prevents the development of mutually beneficial relationships between parties that ensure the accessibility and usability of the correct resources (e.g., data) at the right time under agreed-upon conditions. In this paper, we explore whether the User-Managed Access (UMA) extension to OAuth 2.0 can function as an authorization framework in a decentralized Web environment like Solid, and to which extent it achieves the desired combination of semantic interoperability, data reuse, and Digital Trust. To evaluate this, we identify several requirements. The trust of resource owners (ROs) in requesting parties (RPs) is founded upon several characteristics of a strong authorization system, which governs both access control and usage control: compatibility with state of the art authentication methods and policy modeling languages; flexible and understandable authorization requests; delegation of control; and retractable usage control policies. The trust of the RP in the RO, on the other hand, depends heavily on guarantees regarding the context of the transaction, e.g., provenance of data and proof of authorization. We implemented two prototypes combining UMA with the Community Solid Server (CSS, a reference implementation of Solid), and evaluated it with regard to each of these requirements. The first one, which is fully UMA-compliant, already addresses some of the limitations of Solid's default WAC/ACP-based authorization mechanism, but still leaves room for improvement. The second prototype slightly adapts UMA to achieve most of the remaining goals whilst remaining compatible with standard UMA clients. Some required features — notably delegation of control and retractable usage policies — are not specified by UMA, but the specification leaves sufficient room for additional specifications to incorporate them into an UMA server. This underscores the viability of a UMA-based authorization mechanism for the decentralized Web of Data.*

# Introduction

**Access control** systems provide us with **guarantees** about who can access what, making us **confident** that only the right parties can access only the right data. They form the basis of a mutual relationship of **Digital Trust** between the parties exchanging data or services. The plethora of non-interoperable systems currently available to protect resources on the Web, however, make Digital Trust a cumbersome effort: developing and interacting with all those different systems is a costly affair, especially since the goal of the interaction is the data or service itself, not the access control around it. Access control systems are thus a prime example of technology that, in helping us, *works against us*.

How can we make it work *for* us instead, and **make Digital Trust more scalable**? Semantic Web technologies like *Internationalized Resource Identifiers*[9] (IRIs) and the *Resource Description Framework*[10] (RDF) enable humans and machines alike to communicate meaningfully about anything with anyone. They have already delivered a widespread *semantic interoperability*, *automation*, and *reuse* of **public** data and services, through standardized vocabularies, digital identifiers, and the publication of ever-more open data on the Web. However, these technologies are mostly aimed at the interoperability of the resources[11] themselves, rather than the interfaces and protocols through which they are accessed. **Personal data**, and other forms of **private resources**, have therefore largely missed out on these benefits, since they are often stuck behind a non-interoperable wall of access control.

What if we can employ the same technologies to create a more interoperable system protecting private resources? Not only does this make Digital Trust more scalable; it also **makes trust flow more smoothly** between the parties exchanging resources. A shared understanding of which entity provides which data or service under which conditions — and how to access those resources — builds trustful relations, and increases confidence in the provided resources:[12]

- *By relying on a single source of truth, rather than a copy, trust in its up-to-dateness increases.*
- *When the data is provided by an authoritative entity, trust in its correctness increases.*
- *When access is coupled to clear conditions, trust in its correct usage increases.*
- *If one can choose freely between multiple interoperable systems, trust in security increases.*

In this white paper, we will therefore take a first step towards an authorization framework that bridges the gap between access control and the Web of Data, in order to achieve an increased and scalable Digital Trust. In particular, we will look into the ***User-Managed Access***[13] (UMA) extension to **OAuth 2.0**[14], and explore its usage as an access control mechanism for **Solid**[15], a project aimed at the decentralized storage and exchange of personal data. We start by providing some background and terminology about the aforementioned technologies, before specifying the requirements we expect to find in an ideal solution. We then describe our research implementation, and evaluate its possibilities and limitations.

---

[9] Cf. IETF's RFC 3987
[10] Cf. W3C's RDF 1.1 Recommendation
[11] While most of the examples in this paper will be about (personal) data, the requirements, design, and conclusions hold for any kind of resources on the Web.
[12] To read more about how scalable Digital Trust impacts relationships over time, see R. Verborgh's Trust takes time (2024-10-15).
[13] We use both the UMA 2.0 Grant for OAuth 2.0 Authorization specification and the complementary Federated Authorization for UMA 2.0, both by the Kantara Initiative, and refer to both with the term 'UMA', unless otherwise specified.
[14] Cf. IETF's RFC 6749
[15] Cf. https://solidproject.org

# Background

Before diving into the requirements we expect our envisioned system to meet, we give a brief overview of the state of the art regarding the technologies involved. We introduce the Solid Project, its current authorization systems, and their limits; and explain the evolution of OAuth 2.0 and its UMA extension. Throughout this section, we include a number of concepts defined by those specifications. An overview of this terminology can be found in Appendix A.

## The Solid Project

Solid aims to standardize specifications around the *decentralized storage* of data, the access to that data using *digital identities*, and the reuse of that data across multiple applications. Addressing the limitations of traditional Web storage (cf. Figure 1), it envisions a Web where data can operate independent of applications, through technical specifications[16] which detail how to access private data across multiple autonomous locations, called (Solid) *pods*, by proving one's identity (cf. Figure 2). In fall 2024, stewardship of the project was transferred to the Open Data Institute[17] (ODI), a non-profit company dedicated to advancing trust in data,[18] and some of the specifications were taken up by the W3C's *Linked Web Storage*[19] (LWS) Working Group. Solid relies on the power of the Semantic Web to enable a global interoperability between all parties of its decentralized ecosystem.

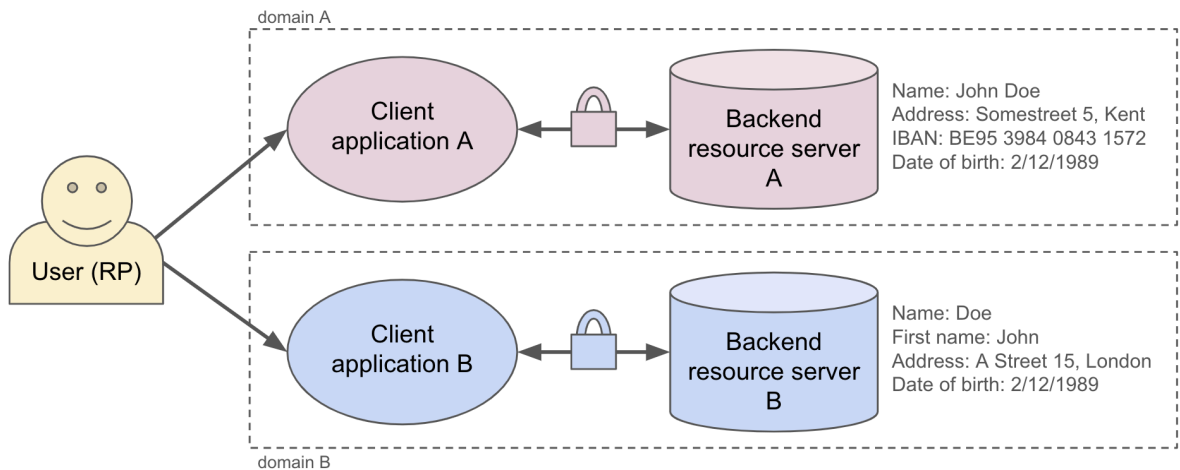

*Figure 1. Traditional Web storage is a domain-specific concern: each corporation hosts its own silo of user data, which it makes accessible through a limited set of (its own) client applications, relying on a user-specific client credential that only works with their service. Clients and servers of different domains are not compatible, and there is a high level of duplicate or even incorrect/out-of-date data. The user themselves has often no direct control over the stored data, and sometimes even no access.*

---

[16] Cf. W3C Solid CG's Technical Reports, in particular it's Solid Protocol Draft CG Report (2024-05-12).
[17] https://theodi.org
[18] Cf. https://theodi.org/insights/projects/odi-and-solid-building-a-future-where-data-works-for-everyone/
[19] Cf. https://www.w3.org/groups/wg/lws

Currently, access control on data in Solid is specified in the [Web Access Control](#) (WAC)[20] specification, and its more recent alternative, the [Access Control Policy](#) (ACP)[21] language. Both specifications work similarly: they provide a language for writing meta-resources, called *access control resources* (ACRs), containing *access control lists* (ACLs) of access control rules (also called *policies*), and an algorithm for finding out which one protects a given resource. While ACP delivers a greater flexibility and extensibility, both specifications still suffer from some limitations.

A. Both specifications lack a **[separation of concerns](#)** between [resource servers](#) (RSs) and [authorization servers](#) (ASs) (cf. infra). What is ubiquitous in modern access control mechanisms, like OAuth 2.0, was not yet so widespread when the initial drafts of WAC, and later ACP, were started. Solid servers therefore incorporate the access control themselves, leading to a request-efficient but **inflexible** system. Since the protection domain is limited to the Solid RS itself, access control management is tailored to the specific interface(s) that this RS provides, and can have no influence on other RSs. In particular, this means that it is impractical to manage and audit access control over multiple servers from a single place, or use the same policies to govern multiple domains. Moreover, information regarding the digital identity of the user is unnecessarily revealed to the RS.

   An additional issue that arises from this — but is not WAC- or ACP-specific — is that it **limits available [authentication methods](#)**. While both [policy modeling languages](#) support extending policy conditions beyond the standard included few, the technical support for those extensions is tightly bound to the evolution of the Solid RS. Currently, this means that one can only authenticate by exposing the full [WebID](#)[22] of the user via [OpenID Connect](#)[23] (OIDC). The options are thus limited to two methods that reveal much about the identity of the entity accessing a resource. Adding support for other authentication methods would require a change in the specification and implementation of the Solid RS. On the contrary, if support for authentication methods was decoupled from the RS, and/or handled by some negotiation mechanism, an AS could implement the extensions it wants to support, while the RS would not have to implement any change at all.

B. The modeling of policies forms the basis of the authorization capabilities of any protected system. Within WAC and ACP, the (client–server) interaction protocol is tightly integrated with both the policy language and the algorithms to calculate authorization decisions. Hence, both specifications restrict the flexibility of the decision making. In previous work, we showed how difficult it is to combine WAC or ACP with more complex policy modeling languages like the **[Open Digital Rights Language](#)** (ODRL)[24], the current state of the art in modeling legal and user requirements for data protection in open Web ecosystems.[25] For example, ODRL can easily incorporate machine-readable expressions of legislative requirements,[26] using ontologies such as the [Data Privacy Vocabulary](#)[27] (DPV).[28]

---

[20] Cf. W3C Solid CG's [WAC Editor's Draft](#) (2024-05-12)
[21] Cf. W3C Solid CG's [ACP Editor's Draft](#) (2022-05-18)
[22] Cf. W3C WebID CG's [Draft Community Report](#) (2014-05-12)
[23] Cf. OpenID's [OpenID Connect Core 1.0](#)
[24] Cf. W3C's [ODRL Information Model 2.2 Recommendation](#)
[25] W. Slabbinck, et al. *Enforcing Usage Control Policies in Solid using Rule-Based Web Agents*. 2nd Solid Symposium. Leuven, 2-3 May 2024. Accepted for publication. ([preprint](#))
[26] B. Esteves, et al. *ODRL profile for expressing consent through granular access control policies in Solid*. *2021 IEEE European Symposium on Security and Privacy Workshops (EuroS&PW)*. IEEE, 2021.
[27] Cf. W3C Data Privacy Vocabularies and Controls CG's [DPV 2.0 Final Community Group Report](#)
[28] H. J. Pandit, et al. *Data Privacy Vocabulary (DPV) – Version 2.0.* 23rd International Semantic Web Conference 2024. Baltimore, US, 11-15 November 2024. Accepted for publication. ([preprint](#))

C. Access control rules (policies) are **targeted at resources in a hierarchical way**. In previous work, we demonstrate how this approach assumes a symmetry in the reading and writing of resources, which prevents a full independence of data and application. Because it requires applications to have knowledge about the organization and permissions of data, this leads to a proliferation of non-interoperable application-specific APIs on top of Solid.[29]

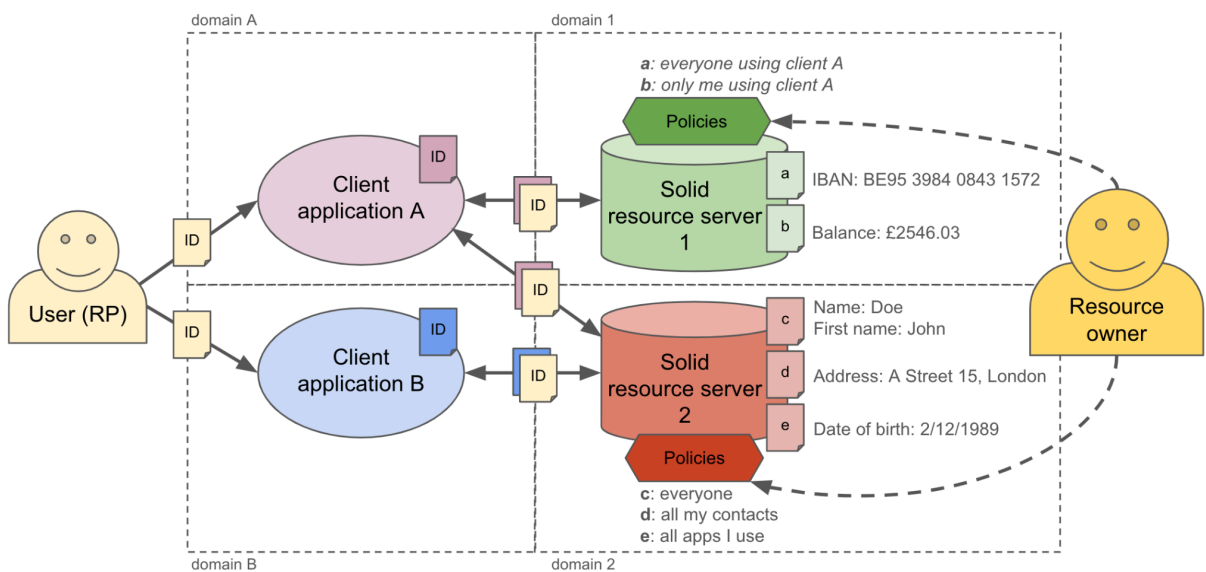

**Figure 2.** *Decentralized Web storage as envisioned by Solid separates the data from specific applications. Data is stored in the most relevant place, reducing duplication and out-of-dateness. Users can access data on each RS through any Solid-compliant application, provided the RO has set the right policies on the RS in question (dashed lines). Policies are set per resource, directly on the RS where that resource is stored. Users identify themselves using OIDC-secured WebIDs, which are passed through the client application to the RS.*

# OAuth 2.0 & UMA

The idea for OAuth saw the light in the early 2000s, when developers of the *read-write Web* reached the limits of what was possible with pure *HTTP Authorization*[30]. Lacking an elaborate authorization system — which was an explicit *non-requirement* in the original design of the Web[31] — multiple companies were creating their own proprietary, mutually non-interoperable solutions. However, since this threatened to lock everyone — companies and users alike –- in a technologically suboptimal dependency on one specific system, a group of organizations (notably Twitter, Google and Flickr) sat together to come up with an open interoperable specification.

One specific desirable feature was a secure and flexible way to delegate access from a user to a particular *client application*. After a lukewarm reception of the initial version, the specification was adopted by the IETF and improved to version 2.0[32], which quickly became the de facto standard of access control on the Web. Major factors contributing to the popularity of OAuth 2.0 were its increased flexibility, performance and scalability. One of the key decisions which achieved this is a **separation of concerns** between the RS and the AS: while the former manages the resources, the latter manages the access control. An opaque *token* from the AS signifies access granted to the client. Neither server needs to know anything else about the other's internal logic or data; the granularity of control is limited to agreed-upon *scope names*. This additional decoupling is not a mere technical decentralization; it allows different parties to take different responsibilities in the Web of Data, and build specialized trust relations with each other and with end-users.

---

[29] R. Dedecker, et al. *What's in a Pod? A knowledge graph interpretation for the Solid ecosystem.* Proceedings of the QuWeDa 2022, vol. 3279. CEUR, 2022, pp. 81–96. (preprint)
[30] Cf. IETF's RFC 9110
[31] Cf. T. Berners-Lee. *Information Management: A Proposal.* CERN, 1989. (digitized)
[32] Note that the IETF already has a draft for OAuth 2.1, which mostly bundles a number of common OAuth 2.0 security practices.

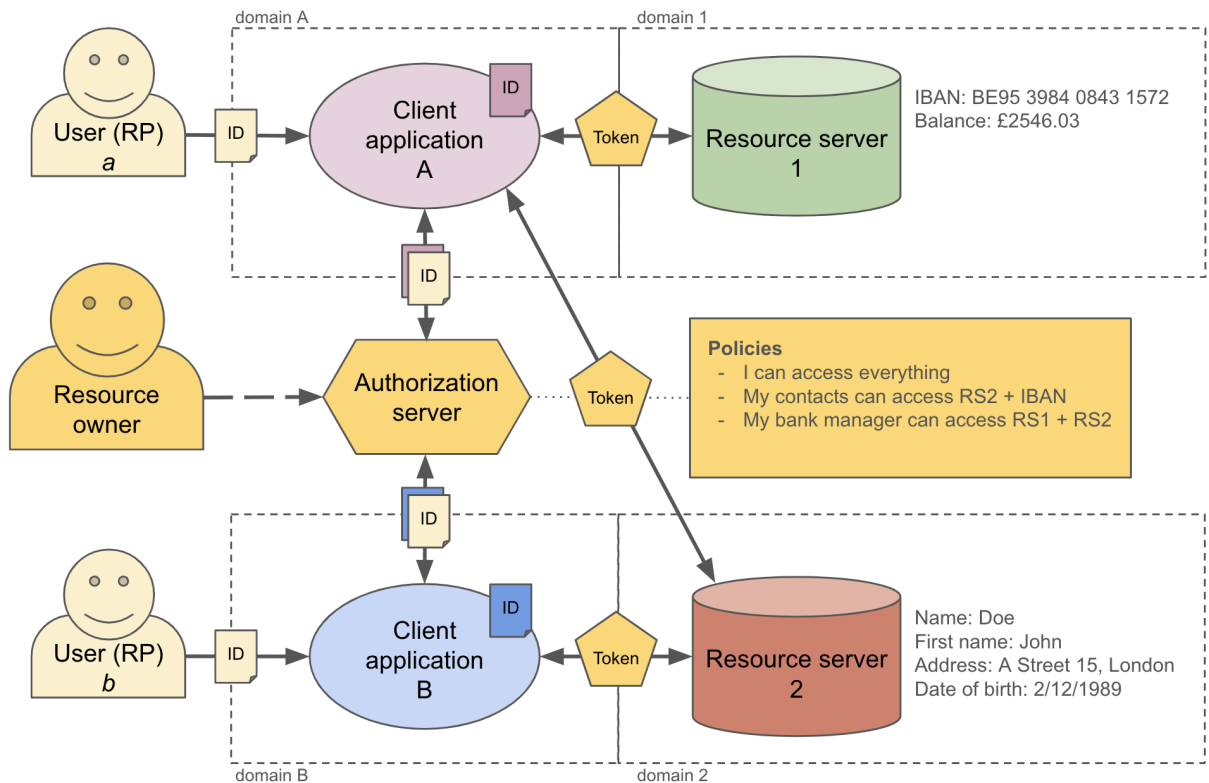

*Figure 3. OAuth 2.0 splits off access control into a dedicated AS, while its UMA extension allows this AS to govern multiple RSs. Data can be stored in the most relevant place, and can be accessed with any UMA-compliant application, provided the RO has set the right policies on the AS (dashed line). Users identify themselves using any authentication method recognized by the AS, upon which the latter generates an opaque token that enables access to the RS. Policies are set per resource **set**, which can be added dynamically to the AS by each RS.*

The core OAuth 2.0 standard only specifies the interfaces and protocol sequences for communication between the client and both servers. This leaves a lot of freedom to the implementer, but numerous formal extensions have been specified. Specifically, UMA extends OAuth 2.0 with the following:

1. It separates the roles of *requesting party* (RP) and *resource owner* (RO),[33] thereby allowing **delegation to other people and organizations**, rather than only to applications. This enables multi-user resource sharing scenarios, in which access policies changed by one user can asynchronously affect the access to resources of another user.

---

[33] From a legal perspective, the concept of 'ownership' over data/resources is notoriously difficult, up to a point that it merits avoidance of the term. Here, and throughout this paper, however, we employ it in its well-defined technical sense: an entity capable of granting access to a protected resource.

2. It specifies an interaction between RS and AS to register new **resource sets**, each with their own characteristics (e.g., scopes). Upon a failed attempt to access a resource, clients receive a *permission ticket* from the RS, which indicates to the AS which resource set and scopes would cover the desired access. This design enables an AS to dynamically protect multiple RSs in different protection domains, allowing ROs to easily manage their resources through an AS of their choice.

3. Using these tickets, it also supports a more **dynamic negotiation** between clients and the AS, allowing the latter to request additional authentication proofs depending on the requested access. Figure 3 gives an overview of the resulting architecture. By leveraging the advantages of OAuth 2.0, and in particular its UMA extension, we hope to improve upon the current access control mechanisms in Solid, thereby increasing its interoperability and level of automation, in order to achieve an increased and scalable Digital Trust. In the next section, we will make these requirements explicit.

# Requirements

As introduced above, the purpose of this paper is to discover to what extent the combination of a modern access control mechanism like OAuth 2.0's UMA with a decentralized Web framework like Solid can increase Digital Trust. In this section, we formulate the requirements by which we will assess the achievement of that purpose.

Digital Trust forms part of a shared relationship of **mutual trust** between the parties exchanging data or services, and is upheld by the trust of both parties in a digital system that provides certain guarantees. While other factors might contribute to their shared relationship, an increase in Digital Trust leads to an increase in their mutual trust. The guarantees ROs and RPs require of each other and of the system differ, and sometimes oppose each other, requiring a fine balance. In our assessment, we include requirements that aim to strengthen Digital Trust in both directions.

## Access & usage control

The trust of an RO in the RP accessing and, consequently, using a resource, is largely founded upon the ability of the former to control who can access what and under which conditions (e.g., for which purpose). In other words, the RO wants the right RPs to have access to the right resources for the right reasons. A system's features for regulating this are collectively known as ***usage control*** (which typically includes *access control* under its umbrella). We distinguish a number of requirements that increase this level of control.

   I.  **Secure and granular authentication methods.** The RO wants to know exactly *who* is accessing and using their resources. Trust increases when this *identity* is guaranteed to be correct through some authentication procedure, which possibly involves a trusted third party. However, while for the RO the strength of the guarantee is what matters most, this is often counter-balanced by the desire of the RP to reveal as little as possible about themselves, i.e., not more than is necessary to gain access to the resource. To accommodate all parties in a decentralized ecosystem, with different use cases and varying levels of sensitivity, a ***flexible and extensible authentication mechanism*** is therefore required, which allows easy integration with both existing and future state of the art authentication methods, yet ensures **privacy-minded need-to-know information sharing**.

   II. **Expressive user-managed policies.** At the base of access and usage control lies the ability to set and change the *usage control policy* (UCP) rules that govern who can access what and under which conditions. The expressive power of UCP rules, and thereby the granularity of control, is determined by the detail with which those three aspects can be described: the party that is granted access, the conditions under which this access is granted, and the resources to which the policy applies. The UCP **modeling language must be sufficiently expressive** on all three fronts.[34] Preferably, the choice of modeling language is also fully orthogonal to the access control protocol itself, so that both can **evolve independently**.

   Increased granularity of modeling is only useful in as much as it adds semantics that help the RO in making useful distinctions in one of the three mentioned aspects. One such distinction is the classification of resources into ***resource types***. The ability to set policies by resource type greatly improves their usefulness: both parties can be sure that the access and usage conditions of such a policy hold for all resources of that type, and only those resources, possibly extended to future additions of resources of that type. Because of their powerful function, we require an authorization system to handle such policies.[35]

---

[34] For a more in-depth discussion on how the granularity of interfaces, policies and queries intertwine, see R. Verborgh's *The Web's data triad* (2024-5-30).

III. **Expressive authorization requests.** Apart from providing authentication, the RP must also express which resource (type) it wants to access, and under what conditions (e.g., for which purpose). The more *specific* this information can be communicated, and the more *understandable* it can be presented to the RO, the more trust the latter can place in providing the resource to the RP. A system facilitating trust must therefore allow for *expressive authorization requests*, possibly reusing (part of) the policy modeling language.

IV. **Delegation of control.** Many non-trivial use cases of access and usage control, especially in decentralized environments, involve **multiple parties having (partial) control** over the resource, i.e., each of these parties can to a higher or lower extent decide certain aspects of the UCPs. To accommodate these use cases, an authorization system should include some mechanism for [delegating (partial) control](#) between parties, and governing this delegation, possibly changing or retracting it. Note that this requirement is about actual control, i.e., direct influence on policies; the fact that many credentials can be passed around, allowing parties to mask as another, goes against the aim for more Digital Trust.

V. **Retractability of policies.** Just like delegated control, policies themselves should also be [retractable](#). In most authorization systems, this is easily done with regard to mere access control. However, *retractability of more general usage control policies* remains a challenge. While this field is still in its infancy, we ultimately want a system in which parties can agree on policies that are subject to change. Ideally, our proposed system should therefore at least leave open the possibility of adding such mechanisms.

## Transaction contextualization

In the opposite direction, trust of an RP in (the resource provided by) an RO can largely be established by certain **guarantees regarding the [context](#)** of the transaction. This context includes several different claims that can be made either by the RO themselves or by third parties with whom there exists a pre-established trust relation.[36]

VI. **Provenance of the resource.** A resource without a known [provenance](#) is like a bank check without a signature: worthless. Any data can be generated, any service emulated. Not knowing where the following statements come from, they all hold equally little value: "Ruben is born 27/2/1985", "Ruben is born 28/2/1986", "Ruben is born 29/2/1987". It is its origin with the population registry, and one's trust in that registry, that makes one of them special.

---

[35] The need for resource types in more complex policy modeling also introduces another benefit the AS might provide. As a trusted system that knows about an ROs resource types, it can automate the [discovery of resources](#): in negotiation, an RP can then ask an AS whether certain (types of) resources are available and, if so, request access to them; based on the relevant policies, the AS then decides whether or not to share this information with the RP. While this feature requires the AS to know which (types of) resources are stored where, it enables a level of automation that lets resources work for us, rather than the other way around.

[36] For a more elaborate discussion on the need for such a contextualization, read all about *trust enveloppes* in R. Verborgh's *[No more raw data](#)* (2023-11-10).

Of course, on the Web, and especially on domains served via an encryption protocol like *HTTPS*, a minimal sense of provenance is always available; but it is limited to a coarse and static indication of who technically provided the resource; not who created it, nor who authorized its access and usage, or who can vouch for its correctness. To increase trust in the resource itself, it is therefore important to model and transfer more **complex forms of provenance**. Some data formats, e.g., [Verifiable Credentials](#)[37] (VCs), are designed for this explicitly, using methods like *digital signatures* to **guarantee the provenance** of data. However, an approach in which such guarantees can be made **independent of the format** of the representation of the resource would result in a much more interoperable ecosystem.

VII. **Auditability.** Getting access to a resource is one thing; proving that one had access to a resource, and that one could indeed use it in a certain way, is a completely different story. For both parties, this ability drastically increases trust in each other. The RO can rest assured that the party using the resource can be [*audited*](#) to make sure the resource is indeed only used within the bounds of the corresponding UCPs. When such an audit takes place, for whatever reason (valid or invalid), a benevolent user of the resource can then rest equally assured that they can prove black on white that they are indeed within their rights. To achieve this, the transaction context must be available to the user, containing the details of the conditions specified in the relevant UCPs.

Note that, similar to the authentication requirement above, auditability introduces a careful balance of *need-to-know information sharing*: while disclosing certain details about UCPs is necessary to provide this [***proof of authorization***](#), revealing the UCPs themselves can introduce a security vulnerability, allowing malicious actors to employ more targeted approaches, tailored to the specific conditions of certain UCPs. This balance must therefore be taken into account when designing both the authorization system and the policy modeling language.

VIII. **Other contextual information.** While provenance and auditability are two extremely important aspects of the transaction context, one can conceive of other pieces of information that might increase Digital Trust between two parties, e.g., *temporal information* about the access and usage of a resource, or of the history of the resource. The format of a transaction context must therefore be sufficiently open to **allow for a variety of information**.

# Facilitating features

Trust of both parties in an authorization system can be increased by certain factors that facilitate its adoption and evolvability. Because the influence of these factors is often indirect, we consider them of secondary priority compared to the requirements in the previous categories.

IX. **Separation of concerns.** The party that is technically managing the policies and making the access control decisions (i.e., the AS) must be separable from the party hosting the resources (i.e., the RS). Not only does this prevent the RS unnecessary access to authentication data; it also achieves a more **thorough decentralization**, allowing a policy system to govern multiple domains of resources. This not enables ROs to swap out any authorization system for one they trust more; it also allows them to more easily manage resources that are spread over disparate RSs, if so desired. For example, a system governing multiple RSs could provide the RO with an overview of resources, policies and delegations, which an RS-bound system cannot, and even apply the same UCP to similar resources on different RSs.

---

[37] Cf. W3C's [Verifiable Credentials Data Model v1.1 Recommendation](#) (and [v2.0 Candidate Rec.](#))

X. **Complex and dynamic negotiation.** Most access control systems employ a one-off exchange of credentials: at most, a general overview of supported authentication methods is available, from which the client chooses one it hopes to fulfill, upon which it sends the necessary credentials according to that method. This is a very restricted model, since it prevents the server from *tailoring the available methods to each particular client*, taking into account the specific policies that would enable that client access. Furthermore, it requires that all necessary credentials can in fact be retrieved and delivered in one go, which often necessitates them to pass through the possible *insecure environment* of the client itself. A more elaborate authentication model would therefore allow a more complex and dynamic interaction between the client and the server.

XI. **Integration with existing technologies.** Both the trust in a system and the adoption of it are also dependent on the ease with which that adoption can take place. People tend to use technologies that are *familiar*, i.e., which they have used before. One immediate implication of this is a preference for a system that integrates easily with the *existing Web architecture*: a solution that follows, combines, or extends popular and well-tested technological standards will have a higher chance of becoming popular and subsequently well-tested itself; and if more of such technologies are compatible with it, it will more easily integrate into already existing products and ecosystems.

XII. **Performance.** While this paper primarily explores technical possibility constraints, it does not hurt to keep the performance of the proposed solution in the back of our minds. Solutions that already perform poorly in theory will definitely be so in practice, and more performant ones will — ceteris paribus — be adopted more easily. For this reason, we should keep an eye out for obvious performance hurdles.

# Prototype

In order to test the extent to which UMA achieves the requirements listed above, we implemented a prototype governing access control to Solid servers. After briefly describing the implementation, we will evaluate the prototype with regards to the requirements. We then look into each limitation, and try to overcome them by implementing several modifications on top of the initial prototype.

## Prototype implementation

We implemented the prototype in TypeScript, to stay in line with most of the existing tooling around Solid, and integrated it with the [Community Solid Server](#)[38] (CSS). The prototype is a team effort by the *Knowledge on Web Scale*[39] (KNoWS) group of *IDLab*, continuing on the prior work of Laurens Debackere[40]. The code is publicly available on our [GitHub repository](#),[41] and is divided into three packages:

- The `CSS` package contains an **extension of the CSS**, which replaces its standard WAC/ACP authorization system with modules that implement the UMA specification. Amongst others, these modules implement the interaction with an UMA AS, the verification of an UMA access token, as well as the necessary setup parameters for launching this adapted Solid RS.

- The `UMA` package contains an implementation of the **AS logic** defined by the UMA specification. We implemented this package on an abstract level, restricting it to the code necessary for the external UMA interface, and an internal interface towards a [policy engine](#) (cf. infra). This abstraction enables the package to be used both in combination with a non-Solid RS, and with multiple interchangeable engines.

- The `UCP` package contains a basic implementation of an ODRL-capable **policy engine**. This engine is called by the AS to evaluate incoming requests and decide on granting access or refusing access. It provides the RS with information about missing claims, and extracts the usage control requirements defined by policies, in order to enable transaction contextualization.

To facilitate the integration, as well as to showcase its possibilities, we made the following design choices.

First of all, since OAuth 2.0 employs a separation of concerns between the RS and the AS, UMA leaves us free to choose the authentication methods and policy modeling language. We implemented support for both OIDC identification, because this is the standard in Solid, and for VCs, because of their popularity as an upcoming standard. For modeling policies, we picked ODRL, because of its flexibility and the traction it is gaining as a standard.

Second, although the separation of concerns means that an RS can choose which AS decides over which scope(s) of which resource(s), for our initial adaptation of the CSS we decided to simply configure a **single AS for all resources**, since this does not influence the interaction model of the protocol itself. This limitation is easily lifted by a future modification that allows a different AS to be configured per pod, or even per resource or scope.

---

[38] Cf. their [GitHub repository](#) or [documentation](#)
[39] Cf. their [web page](#)
[40] Cf. his [GitHub repository](#)
[41] https://github.com/SolidLabResearch/user-managed-access/

Third, with regard to policy evaluation, we made an abstraction of a **policy engine**, to easily switch out implementations. We provided an initial implementation of an **ODRL policy engine**, supporting a limited profile of ODRL capabilities. For example, we can write a policy allowing automatic access to our shoe size to all organizations providing with a VC that they are registered as a shoe seller in the Flemish enterprise registry, but allow only our favorite one — identified by its OIDC token — to send a birthday card — with discount, of course! — to our inbox in a two-week period around our birthday. Additional requirements can easily be implemented by extending this ODRL-based engine, e.g., to model legal rights and obligations using DPV.

Within the above constraints, our research setup allows the following. When a client attempts to access a certain resource in a Solid pod, it is pointed to the corresponding UMA server. The client then requests this server access to the desired resource, interacting in accordance with the UMA specification. If access is granted, the client receives an OAuth 2.0 token. When the client presents this token to the Solid server, the latter will verify its validity, and (if valid) allow access to the resource. Adaptations to this initial setup, based on our evaluation of it concerning the requirements listed above, will be elaborated upon further below.

| Requirement | WAC/ACP | UMA | UMA+ |
|---|---|---|---|
| **I.** Authentication methods | ⚠️(hard to extend) | ✅(unrestricted) | ✅(unrestricted) |
| **II.** Expressive policies | ❌(fixed language) | ✅(unrestricted) | ✅(unrestricted) |
| **III.** Expressive requests | ❌(only URI + scope) | ⚠️(only ticket) | ✅(unrestricted) |
| **IV.** Delegation of control | ❌(all or nothing) | ⚠️(unspecified) | ⚠️(unspecified) |
| **V.** Retractable UCPs | ⚠️(unspecified) | ⚠️(unspecified) | ⚠️(unspecified) |
| **VI.** Provenance | ⚠️(unspecified) | ⚠️(unspecified) | ⚠️(unspecified) |
| **VII.** Auditability | ⚠️(unspecified) | ⚠️(unspecified) | ✅(grant in context) |
| **VIII.** Other context | ⚠️(unspecified) | ⚠️(unspecified) | ⚠️(unspecified) |
| **IX.** Separation of concerns | ❌(no separation) | ✅(separate AS) | ✅(separate AS) |
| **X.** Access negotiation | ❌(one-shot request) | ✅(dynamic) | ✅(dynamic) |
| **XI.** Integration | ❌(custom RDF) | ✅(OAuth + RDF) | ⚠️(idem + mods) |
| **XII.** Performance | ✅(one-shot to RS) | ⚠️(some issues) | ✅(issues fixed) |

*Table 1.* Overview of the evaluation of UMA according to our requirements, including an assessment of WAC/ACP for comparison. The column 'UMA' contains the evaluation of the initial, fully standard-compliant implementation of our prototype, as described in the section Evaluation of Requirements. The evaluation of the prototype including non-compliant modifications, as described in the section Possible Adaptations, is listed under 'UMA+'.

# Evaluation of requirements

We evaluated the implementation described above on each of the requirements we set. An overview of our evaluation can be found in Table 1. In this section, we will discuss them.

By choosing to look into the potential of UMA as an authorization system for the decentralized Web, and in the context of Solid substituting it for the current WAC/ACP approach, we already ticked off a number of requirements presented above. Being an OAuth 2.0 extension, UMA inherently enables a **separation of concerns** (requirement [IX](#)), decoupling both the **authentication method** ([I](#)) and the **policy modeling language** ([II](#)) from the RS. This allows us to freely exchange these aspects of the system without impacting the rest of it, thereby addressing already two of the three issues with WAC/ACP (cf. supra). Concretely, as mentioned earlier, our prototype models policies in ODRL, and accepts authentication with OIDC identity tokens or VC claims.

Moreover, UMA is specifically designed to **govern multiple RSs**, and contains an elaborate authorization request mechanism that allows for **complex and dynamic negotiation** ([X](#)) interactions. For example, our prototype implementation can be configured to answer authorization requests with a demand for specific claims that are missing. Additionally, it returns the usage control requirements defined by the relevant evaluated policies, together with the access token, when an access grant is returned in response to an authorization request. This allows the RP receiving the grant to store it as **proof of authorization** for a possible audit.

Regarding access and usage control, though, some things are still lacking in UMA. Since the specification only describes the interactions between client, RS and AS, and does not prescribe any interaction between multiple ASs, it does not deliver an interoperable mechanism for **delegation of control** out-of-the-box ([IV](#)). The UMA specification is also primarily focussed on access control in the narrow sense, not the broader **usage control**. They therefore lack support for mechanisms to retract more elaborate UCPs ([V](#)). However, neither does UMA specify restrictions that would stand in the way of combining it with separate specifications to achieve these requirements; to the contrary, its separation of concerns actively encourages it.

One case in which UMA does restrict a requirement concerning access and usage control, is in its strict format of authorization requests. UMA increases the decentralization of OAuth 2.0 by decoupling the identifier of a resource set protected by the AS from the IRIs of the actual resources on the protected RSs. Since clients still need to indicate to the AS which resources (and scopes) they want access to, UMA introduced permission tickets to bridge the gap. However, the specification defines the use of such a **permission ticket as a strict requirement**. Clients can only therefore only request access from the AS after they have first attempted to access the resource in question, and received a permission ticket from the RS. This **hampers the performance** ([XII](#)) of clients that already know this information themselves, which could have addressed the AS directly, and prevents the RP from making a more **expressive authorization request** ([III](#)) including, e.g., the resource type, the purpose of their request, or the legal conditions that apply.

UMA does not provide any specific requirements with regard to **transaction contextualization** ([VI](#), [VII](#), [VIII](#)). While the format of an AS's positive response to an authorization request is open, i.e., it can contain additional non-UMA-specific information, the specification does not provide any requirements concerning the generation, format, or verification of context such as provenance and proof of authorization.

Less urgent — but also more easily amended —- shortcomings are some of the facilitating features we desire of a modern authorization system in a decentralized Web. Most importantly, the UMA specification has two surprising peculiarities that impact its performance (XII) and — in an underhand way — ignore the separation of concerns (IX), making a truly decentralized management of one's UCPs impossible.

First of all, like the rest of OAuth 2.0, UMA is designed with **only private resources** in mind: public resources are typically not registered as protected resources at the AS. However, this **ignores the separation of concerns** (IX), since both AS and RS are then jointly responsible for access control. This is not just a theoretical problem: it has a practical impact on the ability to manage (public and private) resources. After all, UCPs might change: private resources can become public, and vice versa, yet neither RS nor AS has enough oversight to be responsible for their coordination.

A naive solution would be to register all resources at the AS, and let a trivial policy make the desired resources *de facto* public. However, this would entail that access to public resources would also require a round trip to the AS, which would not only constitute a serious **overhead** (XII) but also prevent not-UMA-aware clients from accessing those public resources, thus **complicating integration** (XI). For example, if certain images that are publicly stored in such a pod would be included in an HTML page, a browser displaying that page would not be able to access those images; and similarly for protocols relying on publicly available configuration profiles, like OIDC. In the following section, we propose a way to circumvent this conundrum.

Secondly, with regard to integration (XI), the use of an OAuth 2.0 extension is clearly a choice grounded in a popular and robust ecosystem. If there is one particular lack of compatibility that warrants highlighting, it is the **limited choice of security** on communication between the RS and AS: being an OAuth 2.0 extension, the UMA specification relies solely on a *token-based* approach. While this is great for delegation to front-end applications (e.g., a policy management interface), the API between RS and AS could more easily be protected using any of the popular *key-based* security mechanisms, such as JSON Web Signatures[42] (JWS), Mutual TLS[43] (mTLS), or HTTP Message Signatures[44].

Lastly, on a minor note, all messages between client, RS and AS in UMA are encoded as JSON[45] objects, except for the body of the authorization request, which is URL-encoded to adhere to the specification of the token endpoint by OAuth 2.0. This **lack of a uniform JSON encoding** decreases the developer experience, however: RDF data can easily be communicated as JSON-LD, while there is no clear conversion to URL-encoded strings. The ability to retain the full RDF contents as JSON-LD throughout the whole system would increase the interpretability of these exchanges. On its own, this doesn't weigh up against maintaining compatibility with OAuth 2.0, but taken together with some of the other shortcomings presented above, it might add to the consideration to look into other approaches.

## Possible adaptations

In the previous section, we discussed the strengths and limits of UMA with regard to the requirements we set earlier. While UMA addresses a number of issues with the current WAC/ACP authorization in Solid, there are still some points on which it could improve. In this section, we will therefore look into different ways to address the shortcomings.

One of the more important issues we found was the lack of **integration with public resources**. To address this, we introduced a small variation in the interaction between the AS and the RS: upon the creation of a ticket, the former will hint to the latter whether a resource is publicly accessible, by responding with a 200 status instead of a 201. Based on this hint, the RS can then immediately allow the attempted access, saving the client a round-trip to the AS, and making access to public resources also viable for clients that are not aware of the UMA protocol.

---

[42] JSON Web Signatures, cf. IETF's RFC 7515
[43] The use of Transport Layer Security (cf. IETF's RFC 8446) to authenticate both server *and* client.
[44] Cf. IETF's RFC 9421
[45] Cf. IETF's RFC 8259

To improve the performance for RPs that know which permission to request, we also added a shortcut so that they can go directly to the AS of the RO, *without a ticket*. A new ticket is then automatically generated, targeting one or more resources matching the request but skipping the first part of the UMA authorization flow. Note that we are not the first to do so: Keycloak[46], a popular open source IAM[47] solution, made a similar modification in their implementation of UMA.[48]

Another (minor) issue we addressed in the interaction between RS and AS, is the limited security options on their communication. To ensure their mutual trust, we chose to employ a direct **key-based security** approach using *HTTP Message Signatures*[49] with a [JSON Web Key Set](#)[50] (JWKS) on a [Well-Known URI](#)[51], rather than the more involved token-based security specified by UMA, which is more suited for interaction with front-end interfaces.

The changes above involve breaking deviations in the protocol regarding the interaction between RS and AS, but form a **clean extension on the interfaces towards the client**. They increase integration and performance, whilst remaining compatible with standard-conforming UMA clients. However, deviating from the standard breaks integration with the existing ecosystem around UMA and OAuth 2.0, which might hamper adoption. This choice should therefore only be made after careful consideration of the benefits.

An alternative approach is to find another specification that corresponds better with our requirements. One such candidate is the [**Grant Negotiation and Authorization Protocol**](#)[52] (GNAP). While it is a new protocol that is not compatible with OAuth 2.0, it is heavily based on it, and includes parallels to the OAuth 2.0 core as well as OIDC, UMA, and some other extensions, such as [**Rich Authorization Requests**](#)[53]. It also uniformly employs openly extensible messages in JSON, and allows for a wide range of server-to-server security options.

A few requirements remain absent, both in our adapted UMA approach and in GNAP. These include the abilities to delegate control to another party, and to retract UCPs that go beyond mere access control. However, given that both UMA and GNAP are designed with separation of concerns in mind, we are confident that both requirements can be incorporated into both architectures.

---

[46] Cf. https://www.keycloak.org
[47] Identity and Access Management
[48] Cf. Keycloak's documentation on its [Authorization Services, § 8](#)
[49] Cf. IETF's [RFC 9421](#)
[50] Cf. IETF's [RFC 7517](#)
[51] Cf. IETF's [RFC 8615](#)
[52] Cf. IETF's [RFC 9635](#)
[53] Cf. IETF's [RFC 9396](#)

# Conclusion

Access control forms the foothold for Digital Trust between parties exchanging data or services but, lacking interoperability, it is often a cumbersome practice, and makes private resources miss out on the automation and reusability found on the open Web. Using the OAuth 2.0 User-Managed Access (UMA) extension as a decentralized authorization framework for the Web of Data, we overcome some of the obstacles encountered with existing approaches, in order to make Digital Trust more scalable.

To assess our proposal, we identified several requirements. For the trust of resource owners (ROs) in requesting parties (RPs), we include several characteristics of a strong **access and usage control** mechanism: *(I)* secure yet privacy-minded authentication methods, *(II)* an expressive and independently evolving policy modeling language, *(III)* specific and understandable authorization requests, *(IV)* the capability of delegation of control, and *(V)* the ability to retract usage control policies (UCPs). To increase the trust of the RP in the RO, we focus on **guarantees regarding the context** of the transactions, including *(VI)* complex forms of provenance, independent of the resource format and *(VII)* secure and auditable proofs of authorization, *(VIII)* all in an open and extensible format. We also propose a number of **facilitating characteristics** for the digital system: *(IX)* a separation of concerns between resource server (AS) and authorization server (RS), *(X)* support for a complex and dynamic negotiation between client and AS, *(XI)* integration with existing technologies, and *(XII)* the overall performance of the system.

To test the extent to which UMA achieves these requirements, we implemented a prototype integrating UMA with the Community Solid Server (CSS), and evaluated it on each requirement. UMA surpasses Solid's default WAC/ACP-based authorization mechanism in its flexible authentication methods, decoupled policy modeling language, and support for complex/dynamic negotiation and auditing. However, it still leaves space for improvement. Some important features are not specified in UMA, including the mechanisms for **delegation of control**, **retraction of UCPs**, and **contextualization of transaction**. Because of the design of UMA (and OAuth 2.0), however, such mechanisms could be incorporated into an UMA server based on additional specifications.

We also found some requirements that are less easily met by UMA, since they are in conflict with certain aspects of the specification: UMA's **authorization request format** allows little freedom, and the protocol's design introduces some **performance** hurdles. Moreover, while UMA is well-positioned in the OAuth 2.0 ecosystem, a number of its choices around **message format** and **security options** might make it less attractive to developers. To get an idea of how much the UMA specification would have to be changed in order to address these shortcomings, we extended our prototype until it checked most of these requirements. While these adaptations involve breaking deviations with regard to the interaction between RS and AS, we were able to remain compatible with standard UMA clients. This leaves us optimistic about the perspectives of a UMA-based authorization mechanism for the decentralized Web of Data.

Further research includes, in the long term, delegation of control and the retraction of UCPs. In the short term, it would be interesting to expand our prototype to allow different ASs per pod or per resource, and to further integrate the policy engine with DPV, as well as with the latest work on ODRL Formal Semantics[54]. Moreover, while extending existing specifications in a backward-compatible way is a standard practice in the OAuth 2.0 ecosystem, the more we deviate from the standard, the more we lose integration with the existing support for UMA, which might hamper adoption. We will therefore also compare our current work with other specifications, such as the **Grant Negotiation and Authorization Protocol** (GNAP), which might achieve more requirements without deviating from the standard.

---

[54] Cf. W3C ODRL CG's [ODRL Formal Semantics Draft CG Report](#) (2024-10-25)

# Acknowledgments

This research was funded by SolidLab Vlaanderen (Flemish Government, EWI and RRF project VV023/10) and by the imec.icon project PACSOI (HBC.2023.0752), which was co-financed by imec and VLAIO and brings together the following partners: FAQIR Foundation, FAQIR Institute, MoveUP, Byteflies, AContrario, and Ghent University – IDLab.

# Appendix A: Terminology & abbreviations

**Access control**: The process of managing the authorization of parties to access resources.

**Agent**: Any entity capable of acting on its own behalf; e.g., a person or organization.

**Authorization**: Permission to access and/or use certain resources; or the process of granting it.

**Access Control Policy (ACP)**: A specification defining a policy modeling language, and an algorithm to resolve authorization requests based on policies in this language; cf. W3C Solid CG's [ACP Editor's Draft](#).

**Auditability**: A characteristic attributed to a system involving transactions when it enables those transactions to be verified in a systematic and reliable manner. In the context of this paper, the transactions in question are authorizations granted to parties to access and/or use certain resources.

**Authentication**: A party's proof of one of their (partially) identifying characteristics; e.g., one's age, presence in some registry, or possession of some secret; or the process of providing such a proof.

**Authentication method**: The manner in which authentication (proof) is obtained and/or delivered; e.g., as a VC or through OIDC.

**Authorization server/system (AS)**: The system responsible for protecting resources on an RS, by handling authorization requests of clients based on policies set by the RO. In the context of this paper, the AS is defined by OAuth 2.0 and UMA.

**Client**: An application interacting with a server on behalf of an agent. In the context of this paper, the client is an application trying to access protected resources, as defined in OAuth 2.0 and UMA.

**Community Solid Server (CSS)**: An open-source reference implementation of the Solid specifications; cf. their [GitHub page](#) or [documentation](#).

**Delegation of access**: Any process by which an agent/user who has access to a resource can let a client application employ that access, either synchronously (e.g., by entering credentials when prompted) or asynchronously (e.g., by providing the application with certain credentials up-front).

**Delegation of control**: Any process by which an agent who has control over a resource, i.e., who can influence UCPs on that resource, can (partially) share that control with another agent, who can then influence those UCPs independently.

**Data Privacy Vocabulary (DPV)**: A machine-readable ontology of concepts for the interoperable representation and exchange of information about the use and processing of personal data based on legislative requirements ; cf. W3C Data Privacy Vocabularies and Controls CG's [DPV 2.0 Final CG Report](#).

**Grant Negotiation and Authorization Protocol (GNAP)**: An upcoming authorization mechanism, taking inspiration from OAuth 2.0, OIDC, and many extensions and best practices; cf [RFC 9635](#).

**HTTP Message Signatures**: A mechanism that guarantees the integrity and authenticity of HTTP communication using digital signatures over HTTP message components, as specified in [RFC 9421](#).

**Internationalized Resource Identifier (IRI)**: A sequence of Unicode characters that identifies a resource, in accordance with [RFC 3987](#).

**JavaScript Object Notation (JSON)**: A popular data interchange format, as specified in [RFC 8259](#).

**JSON Web Token (JWT)**: A compact, URL-safe encoding of a JSON data structure, which can be signed or encrypted to secure the information transfer between parties, as defined by [RFC 7519](#).

**JSON Web Key (JWK), JSON Web Key Set (JWKS)**: A JSON-serialization of a (set of) cryptographic key(s), as specified in [RFC 7517](#).

**JSON Web Signature (JWS)**: A JSON-serialization of content secured with a digital signature, as specified in [RFC 7515](#).

**JWT-Secured Authorization Request (JAR)**: An extension of OAuth 2.0, allowing authorization requests to be encoded as JWTs, and sent to the AS either by value or by reference, as defined by [RFC 9101](#).

**Key-based security**: Any authentication method that identifies parties by their possession of a cryptographic key; e.g., JWS, mTLS, or HTTP Message Signatures.

**Linked Web Storage (LWS)**: A W3C Working Group aimed at a Web where applications, storage, authentication and authorization are more loosely coupled, building on earlier work on Solid; cf. W3C [LWS WG](#) page.

**Transport Layer Security (TLS)**: A mechanism securing the transport layer channel between client and server using cryptographic keys exchanged in an initial 'handshake', as specified in [RFC 8446](#).

**Mutual TLS (mTLS)**: The use of a TLS communication channel in which not only the server but also the client is authenticated with a cryptographic key; cf. [RFC 8446](#).

**Negotiation**: A (two-way) interaction between parties towards mutual agreement. In the context of this paper, negotiation happens between the client and the AS, concerning the options/requirements for authorization. We call a negotiation process dynamic when the interaction can be longer than a single back-and-forth; we call it complex when it can involve different ways of progressing towards agreement.

**OAuth 2.0**: An authorization framework enabling ROs to govern access to their resources by client applications, as specified in [RFC 6749](#).

**Open Data Institute (ODI)**: A non-profit company dedicated to advancing trust in data; cf. [theodi.org](#).

**Open Digital Rights Language (ODRL)**: A policy modeling language, defined by the W3C [ODRL Information Model](#).

**OpenID Connect (OIDC)**: An authentication framework built on top of the OAuth 2.0 protocol, enabling clients to verify the identity of a user; cf. [OpenID Connect Core 1.0](#).

**Personal data**: Data concerning (individually identifiable) people.

**Pod**: An autonomous location of data storage in Solid/LWS.

**Policy, usage control policy (UCP)**: A rule, written in a policy modeling language, describing the conditions under which a resource can be accessed (i.e., an access control rule/policy) or — broader — can be used (i.e., a usage control rule/policy).

**Policy engine**: A resolution mechanism that, based on a set of policies, makes consistent and unambiguous authorization decisions about authorization requests; e.g., the WAC and ACP resolution algorithms.

**Policy modeling language**: A language that specifies how to write down policies in a machine-readable format, for example, to be interpreted by a policy engine; e.g. WAC, ACP, ODRL.

**Private resource**: Any resource that is not public, i.e., a resource of which the access and/or usage are limited by certain conditions.

**Proof of authorization**: Any artifact, produced by an AS upon granting authorization to a client, which that client can at any later time (e.g. during an audit) present as a guarantee that the authorization in question was indeed granted.

**Provenance**: The origin of a resource; in the broad sense this also includes its processing history.

**Public resource**: Any resource that is not private, i.e., a resource that is accessible and usable without conditions.

**Pushed Authorization Request (PAR)**: An extension of OAuth 2.0, which complements the JAR extension with a new endpoint, to which clients can push more elaborate authorization requests to the AS in a secure way, as defined in [RFC 9126](#).

**Resource Description Framework (RDF)**: A framework for representing information in the Web, consisting of a data model centered around sets of subject–predicate–object statements, and providing different data formats compatible with its model; cf. W3C [RDF 1.1 Concepts and Abstract Syntax](#).

**Requesting party (RP)**: An agent requesting access to a protected resource, through a client application. In the context of this paper, this role is defined by OAuth 2.0 and UMA.

**Resource**: Any entity we can describe or interact with. In the context of this paper, it is often (a piece of) data or a service offered on the Web.

**Resource discovery**: The process of discovering available resources related to some agent, possibly adhering to some additional constraints (e.g., a resource type, usage conditions).

**Resource owner (RO)**: An agent controlling access to a protected resource, i.e., capable of granting or denying access to it. In the context of this paper, this role is defined by OAuth 2.0 and UMA.

**Resource server/system (RS)**: The system responsible for hosting the resources of a RO, and serving these resources to clients, in accordance with the access granted to them by the relevant AS. In the context of this paper, the RS is defined by OAuth 2.0 and UMA.

**Resource set**: A collection of resources treated together for some purpose. In the context of this paper, the term is used during the resource registration protocol of UMA.

**Resource type**: A classification of a resource according to some of its characteristics; e.g., its RDF class or SHACL shape.

**Retractable policy**: Any policy which the agent controlling it can retract at a later point in time, and of which this retraction has an actual impact on the ability of already authorized parties to (continue to) access and/or use the resources in question.

**Rich Authorization Requests (RAR)**: An extension of OAuth 2.0, which enables clients to pass more fine-grained, JSON-encoded information in their authorization requests, as specified by [RFC 9396](#).

**Semantic Web**: An extension of the Web in which information is given well-defined meaning, through technologies like IRIs and RDF.

**Separation of concerns**: A core best practice in software architecture, which dictates that different responsibilities should be handled by different (isolated) components. This results in an increased modularity which improves maintainability, scalability, and extensibility.

**Solid**: A project aimed at the decentralized storage, exchange, and reuse of personal data across multiple applications; cf. [solidproject.org](#).

**Token-based security**: Any authentication method that identifies parties by their possession of a token provided by a (trusted) third party; e.g., OAUth 2.0, OIDC.

**Transaction context**: Any information concerning transaction between the RO and the RP, including parameters of the authorization, and meta-data about the resource in question.

**User-Managed Access (UMA)**: A set of specifications extending OAuth 2.0 with party-to-party delegation, dynamic negotiation, and federation of RSs over ASs.

**Usage control**: The process of managing the authorization of certain parties to use certain resources under certain conditions. Usage control typically includes access control.

**Verifiable Credential (VC)**: An upcoming and versatile way to model, store, and exchange data with the signature, provenance, and timestamp information necessary for the claim verification; cf. W3C [W3C VC Data Model](#).

**Web Access Control (WAC)**: A specification defining a policy modeling language based on access control lists, as well as an algorithm for evaluating authorization requests based on policies in this language; cf. W3C Solid CG's [WAC Draft CG Report](#).

**Web Identity (WebID)**: An HTTP(S) IRI that refers to an agent and dereferences to a document (viz. the WebID Document) that describes this agent, as specified by the W3C WebID CG's [WebID Draft CG Report](#).

**Well-known URI**: URI's to registered "well-known locations", starting with the path prefix "/.well-known/", as prescribed by [RFC 8615](#).

**World Wide Web Consortium (W3C)**: An international non-profit organization that coordinates the development of technical standards for the Web; cf. [w3.org](#).